\documentclass[conference]{IEEEtran}
\IEEEoverridecommandlockouts
\usepackage{cite}
\usepackage{amsmath,amssymb,amsfonts}
\usepackage{amsthm}
\usepackage[lined,linesnumbered,boxed,vlined,ruled]{algorithm2e}
\usepackage{graphicx}
\usepackage{textcomp}
\usepackage{setspace}
\usepackage{xspace}
\usepackage{booktabs}
\usepackage{url}
\usepackage[normalem]{ulem}
\usepackage{enumitem}
\usepackage{balance}
\usepackage{float}
\usepackage{subcaption}
\usepackage{caption}
\usepackage{cleveref}
\usepackage{multirow}
\usepackage[table,xcdraw]{xcolor}
\definecolor{mygrey}{RGB}{230,230,240}
\def\BibTeX{{\rm B\kern-.05em{\sc i\kern-.025em b}\kern-.08em
    T\kern-.1667em\lower.7ex\hbox{E}\kern-.125emX}}
    
\newcommand\revise[1]{{\color{black}#1}}

\newcommand{\sysname}{\textsf{StreamTune}\xspace}
\newtheorem{definition}{Definition}

\SetKwRepeat{Do}{do}{while}

\begin{document}

\title{Learning from the Past: Adaptive Parallelism Tuning for Stream Processing Systems}

\author{\IEEEauthorblockN{Yuxing Han\textsuperscript{1}, Lixiang Chen\textsuperscript{1,2}, Haoyu Wang\textsuperscript{1}, Zhanghao Chen\textsuperscript{1},  Yifan Zhang\textsuperscript{1},  \\Chengcheng Yang\textsuperscript{2}, Kongzhang Hao\textsuperscript{3}, Zhengyi Yang\textsuperscript{3}}
\IEEEauthorblockA{\textsuperscript{1}ByteDance Inc,
\textsuperscript{2}East China Normal University,
\textsuperscript{3}University of New South Wales\\
\textsuperscript{1}\{hanyuxing, chenlixiang.3608, wanghaoyu.0428, chenzhanghao.1997, zhangyifan.5\}@bytedance.com,\\
\textsuperscript{2}ccyang@dase.ecnu.edu.cn,
 \textsuperscript{3}\{k.hao, zhengyi.yang\}@unsw.edu.au}}

\maketitle

\begin{abstract}
Distributed stream processing systems rely on the dataflow model to define and execute streaming jobs, organizing computations as Directed Acyclic Graphs (DAGs) of operators.
Adjusting the parallelism of these operators is crucial to handling fluctuating workloads efficiently while balancing resource usage and processing performance. 
However, existing methods often fail to effectively utilize execution histories or fully exploit DAG structures,
limiting their ability to identify bottlenecks and determine the optimal parallelism.
In this paper, we propose \sysname, a novel approach for adaptive parallelism tuning in stream processing systems.
\sysname incorporates a pre-training and fine-tuning framework that leverages global knowledge from historical execution data for job-specific parallelism tuning.
In the pre-training phase, \sysname clusters the historical data with Graph Edit Distance and pre-trains a Graph Neural Network-based encoder per cluster to capture the correlation between the operator parallelism, DAG structures, and the identified operator-level bottlenecks.
In the online tuning phase, \sysname iteratively refines operator parallelism recommendations using an operator-level bottleneck prediction model enforced with a monotonic constraint, which aligns with the observed 
system performance behavior.
Evaluation results demonstrate that \sysname reduces reconfigurations by up to 29.6\% and parallelism degrees by up to 30.8\% in Apache Flink under a synthetic workload.
In Timely Dataflow, \sysname achieves up to an 83.3\% reduction in parallelism degrees while maintaining comparable processing performance under the Nexmark benchmark, when compared to the state-of-the-art methods.
\end{abstract}


\section{Introduction}

Distributed Stream Processing Systems (DSPSs), such as Apache Flink~\cite{flink}, Timely Dataflow~\cite{2024timely}, Linkedin Samza~\cite{samza} and Amazon Kinesis~\cite{kinesis}, have been widely applied in managing and analyzing real-time data across various industries.
In contrast to traditional batch processing~\cite{hadoop,pig,hive}, which handles large volumes of data at fixed intervals, DSPSs are built for continuous execution, enabling near-instantaneous responses to incoming data.
Notably, ByteDance now operates large-scale Flink clusters with nearly 13 million CPU cores and exabytes of memory, supporting over 100,000 concurrent streaming jobs distributed across 150 countries~\cite{mao2023streamops}. 
These systems rely on the dataflow model~\cite{akidau2015dataflow,murray2013naiad,carbone2015lightweight} as the foundational abstraction for defining and executing the continuous streaming jobs.
This model facilitates scalable
execution by organizing computations into Directed Acyclic Graphs (DAGs).
In this model, nodes represent operators that perform computations on incoming data, while edges capture data dependencies between these operators.
Dataflow execution relies on asynchronous message passing, allowing each operator to process data independently, achieving both high throughput and fault tolerance.
In real-world applications, dataflow execution should accommodate fluctuating workload characteristics, such as varying data arrival rates.

Traditionally, system engineers manage these fluctuations by manually adjusting the parallelism of dataflow operators to match different workload demands. 
This process involves increasing the parallelism (i.e., scaling out) during peak periods to maintain performance and decreasing the parallelism (i.e., scaling in) during off-peak times to conserve resources.
However, manual tuning is labor-intensive and error-prone, which often results in suboptimal resource allocation.
Ineffective adjustments might lead to over-provisioning during periods of low demand, resulting in resource wastage; or under-provisioning during sudden workload peaks, potentially causing violations of Service Level Objectives (SLOs)~\cite{mei2020turbine}. 
As a result, automatic parallelism tuning has emerged as a critical need in DSPSs
to ensure both resource efficiency and system reliability in response to unpredictable workloads~\cite{floratou2017dhalion,kalavri2018three,fu2017drs,lohrmann2015elastic,castro2013integrating}.


\textbf{Existing Methods.} 
Approaches of automatic parallelism tuning in stream processing systems can be broadly classified into two categories: \textit{rule-based} and \textit{model-based}. 
On the one hand, rule-based methods~\cite{floratou2017dhalion,mei2020turbine,castro2013integrating,xu2016stela} utilize predefined thresholds or heuristic-driven rules to trigger tuning actions.
For instance, a common rule would scale up the system when CPU utilization exceeds 70\% for more than five minutes. 
Although these methods are straightforward to implement, they often struggle to capture complex relationships between workload characteristics and system performance, making them less effective in dynamic environments.

On the other hand, model-based methods utilize performance models to guide the parallelism tuning dynamically, addressing the limitations of static rule-based methods.
More specifically, the non-learning-based methods~\cite{kalavri2018three,mei2020turbine} assume a linear relationship between operator parallelism and execution performance, using this to adjust parallelism adaptively.
The learning-based methods~\cite{lian2023conttune,agnihotri2024zerotune} move beyond this linear assumption by leveraging machine learning models to uncover complex, non-linear patterns in workload characteristics and system performance, enabling more precise and adaptive tuning for real-world streaming jobs. 
For instance, ContTune~\cite{lian2023conttune} employs Bayesian Optimization to construct a surrogate model that captures the relationship between parallelism and \textit{operator-level} performance using historical execution data. 
ZeroTune~\cite{agnihotri2024zerotune} combines Graph Neural Networks (GNNs) with zero-shot learning~\cite{pourpanah2022review} to predict \textit{job-level} performance and recommend
{initial}
parallelism degrees for previously unseen streaming jobs.

\textbf{Limitations.} 
Despite recent advancements, existing approaches still face significant challenges that hinder their practical applicability in real-world stream processing systems:

\noindent \textbf{\textit{(C1) Underutilization of Dataflow Execution Histories.}} 
Real-world DSPSs generate extensive execution histories from numerous dataflow runs, providing rich knowledge to enhance parallelism recommendations and speed up the tuning process.
However, existing approaches fail to fully exploit this valuable information.
For instance, ContTune relies solely on the tuning history of the target streaming job, neglecting knowledge from similar past jobs. 
While ZeroTune incorporates past
histories through zero-shot learning, its parallelism recommendation depends on the prediction of the job-level performance metric 
(e.g., latency and throughput).
However, this indirect approach often leads to suboptimal tuning decisions, especially when performance predictions are inaccurate.

\noindent \textbf{\textit{(C2) Insufficient Exploitation of Dataflow DAG Structures.}} 
When executing dataflow DAGs for processing streaming jobs, data flows sequentially from upstream to downstream operators.
Bottlenecks arise when upstream operators process data faster than downstream ones, causing the latter to struggle with excess input.
Effective parallelism tuning should take these inter-operator dependencies into account, which ensures that each operator’s processing ability matches its input rate.  
However, existing methods overlook these structural dependencies, making ineffective tuning decisions. 
While ZeroTune~\cite{agnihotri2024zerotune} employs GNNs to model the structural characteristics of dataflow DAGs, it aggregates operator information to predict job-level performance rather than focusing on operator-level behavior.
This aggregation approach makes it rather hard to identify bottlenecks
at the operator level, thereby limiting its ability to provide optimal parallelism recommendations.




\textbf{Our Approach.} To address the above challenges, we propose \sysname, an adaptive parallelism tuning approach for DSPSs. Our \sysname introduces three key solutions.

\noindent \textbf{\textit{(S1) Pre-training and Fine-tuning Framework:}}
To address C1, \sysname introduces a novel pre-training and fine-tuning framework that leverages global knowledge from historical data for job-specific parallelism tuning.
During the pre-training phase, \sysname learns 
a coarse correlation between parallelism degree and operator-level performance using the historical execution data.
This approach enables effective knowledge transfer across diverse workloads and minimizes the need for extensive retraining when tuning new streaming jobs. 
In the fine-tuning phase, real-time feedback is collected to iteratively refine parallelism recommendations for the target streaming job, with model updates restricted to a lightweight prediction layer to ensure both accuracy and computational efficiency.
To facilitate the parallelism recommendation, parallelism degrees are handled separately from other operator features during \revise{different} phases.

\noindent \textbf{\textit{(S2) Graph Edit Distance-Based Clustering:}}
Also addressing C1, \sysname utilizes a Graph Edit Distance (GED)-based~\cite{gouda2016csi_ged} clustering approach during the pre-training phase to group structurally similar dataflow DAGs in the large-scale historical data.
To improve efficiency and minimize computational overhead, the clustering process leverages graph similarity search techniques~\cite{zheng2014efficient,zhao2013partition} to reduce the number of pairwise distance computations between dataflow DAGs. This approach significantly lowers memory consumption and accelerates the clustering process.
By clustering dataflow DAGs before training, \sysname 
enhances training efficiency by narrowing the training dataset size for each cluster and improves adaptability to diverse workloads during the fine-tuning phase. 
Meanwhile, it could support scalability through independent and parallel pre-training across clusters.


\noindent \textbf{\textit{(S3) Bottleneck Identification and Operator-Level Predictions:}}
%
To address C2, \sysname focuses on identifying and predicting operator-level bottlenecks within the streaming jobs.
This is achieved by systematically analyzing dataflow DAGs to identify bottleneck operators, which are then labeled for training purposes. 
These dataflow DAG with labeled operators are used to train GNNs on a 
operator-level
bottleneck prediction task, enabling the model to capture the structural characteristics of the graph. 
During tuning, candidate parallelism degrees are evaluated based on bottleneck predictions to recommend the optimal configuration.
A key feature of this process is the enforcement of a monotonic constraint on the prediction model.
This constraint ensures that the model aligns with the observed behavior of stream processing systems, where increasing the parallelism of a dataflow operator reduces its likelihood of becoming a bottleneck. By adhering to system dynamics, the monotonic constraint enhances interpretability, making bottleneck predictions and parallelism recommendations both effective and reliable.

\textbf{Benefits of StreamTune.} 
To evaluate the effectiveness and efficiency of \sysname, we deploy it on two representative stream processing systems: Apache Flink and Timely Dataflow.
Experimental results demonstrate that, in Flink, \sysname reduces the number of reconfigurations by up to $\mathbf{29.6\%}$ and achieves a maximum reduction in parallelism degrees by $\mathbf{30.8\%}$, showcasing its resource efficiency under benchmark workloads.
In Timely Dataflow, \sysname reduces parallelism degree by up to $\mathbf{83.3\%}$ while maintaining processing performance comparable to other methods.

In summary, we make the following contributions:
\begin{itemize}
    \item We propose a pre-training and fine-tuning framework that integrates global knowledge with job-specific adaptation, enabling efficient and adaptive parallelism tuning for stream processing systems.
    \item We develop a GNN-based encoder for the dataflow DAGs, trained to predict operator-level bottlenecks. During tuning, the prediction model is enforced with a monotonic constraint to reflect the dynamic behavior of DSPSs.
    \item We introduce a Graph Edit Distance-based clustering approach to group similar dataflow DAGs to improve efficiency and minimize computational overhead.
    \item We validate \sysname by applying it to two representative stream processing systems, demonstrating its effectiveness and efficiency.
\end{itemize}


\section{Preliminaries}

In this section, we first introduce basic concepts in the stream processing systems. Then, we provide a formal definition of the parallelism tuning problem.

\subsection{Stream Processing Systems}

\begin{figure}[tb]
  \centering
  \includegraphics[width=0.95\linewidth]{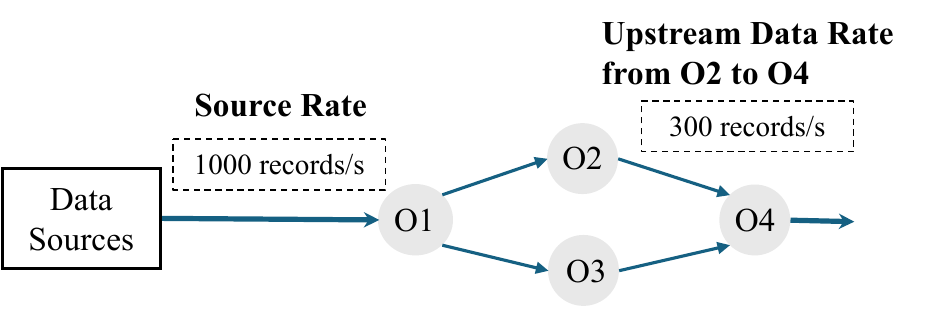}
  \caption{A Logic Dataflow DAG.}
  \vspace{-2em}
  \label{fig:dag}
\end{figure}

\noindent \textbf{Logical Dataflow DAG.} 
%
The \textit{logical} Directed Acyclic Graph (DAG) represents the high-level workflow of operations defined by the application, focusing on task relationships while abstracting execution details. 
Fig.~\ref{fig:dag} illustrates a typical logical dataflow DAG with four operators.
In contrast, the \textit{physical} DAG reflects the actual deployment of operators on computing resources, detailing how the logical plan is partitioned, parallelized, and executed.
In this work, discussions on parallelism tuning pertain to the operators in the \textit{logical} DAG.

\noindent \textbf{Data Sources \& Source Rates.} 
Data sources in a stream processing system refer to the external systems or processes that generate data continuously. These sources include sensors, data feeds, or user inputs.
The source rate specifies the speed at which data is produced. For example, in the dataflow DAG shown in Fig.~\ref{fig:dag}, the source rate is 1000 records per second.
Variations in source rates require adjusting operator parallelism to maintain performance and avoid bottlenecks.


\noindent \textbf{First-Level Downstream Operators.} 
These are the operators in a dataflow DAG that directly receive data from the sources without intermediate operators. In Fig.~\ref{fig:dag}, o1 represents such an operator.
They serve as the initial processing nodes, consuming source data and triggering subsequent computations or transformations in the dataflow.

\noindent \textbf{Upstream Data Rate.} 
The upstream data rate refers to the rate at which an operator receives data from its upstream operators. In Fig~\ref{fig:dag}, the upstream data rate from O2 to O4 is 300 records per second. This rate is dynamic and influenced by the source rate and the processing speeds of upstream operators.

\noindent \textbf{Useful Time.} 
Useful time refers to the total duration an operator actively processes data, excluding idle or waiting periods. It captures the time spent on tasks such as serialization, computation, and deserialization~\cite{kalavri2018three}.

\noindent \textbf{Processing Ability.} 
The processing ability (PA) of a dataflow operator quantifies its data processing rate, measured in records per second over a unit of useful time.
PA depends on factors such as the operator's computational complexity, parallelism setting, and the underlying system configuration.

\noindent \textbf{Backpressure.}
%
In DSPSs, backpressure acts as a flow control mechanism to prevent data overload by slowing data production when an operator cannot process incoming data fast enough. This can trigger a \textit{cascading effect}, where the slowdown propagates throughout the dataflow pipeline, creating performance bottlenecks and degrading overall system efficiency.
\textit{Dataflow-level} backpressure reflects a system-wide issue caused by one or more bottlenecks, while \textit{operator-level} backpressure identifies the specific operators within the dataflow DAG responsible for the performance degradation.

%

\noindent \textbf{Dataflow Execution Histories.} 
A long-running stream processing system accumulates extensive execution histories from various streaming jobs, providing valuable information for analyzing and optimizing future workloads. These records include job-level details (e.g., dataflow DAG \revise{structure}), operator characteristics (e.g., \revise{operator} type, parallelism degrees,
partitioning strategy), and data stream properties (e.g., source rate, tuple width).

\subsection{Parallelism Tuning Problem}
Given a set of streaming job histories $\mathcal{H}$ and a target streaming job with its \revise{\textit{logical}} dataflow DAG, where multiple data sources generate records, each operator must sustain all source rates to maintain optimal throughput.
The objective of parallelism tuning is to use $\mathcal{H}$ to determine the minimum parallelism degrees $\{p_1, p_2,..., p_N\}$ for each dataflow operator, ensuring backpressure-free execution of the dataflow.

\begin{figure*}[h]
  \centering
  \includegraphics[width=1\linewidth, trim=0cm 0.22cm 0cm 0.22cm, clip]{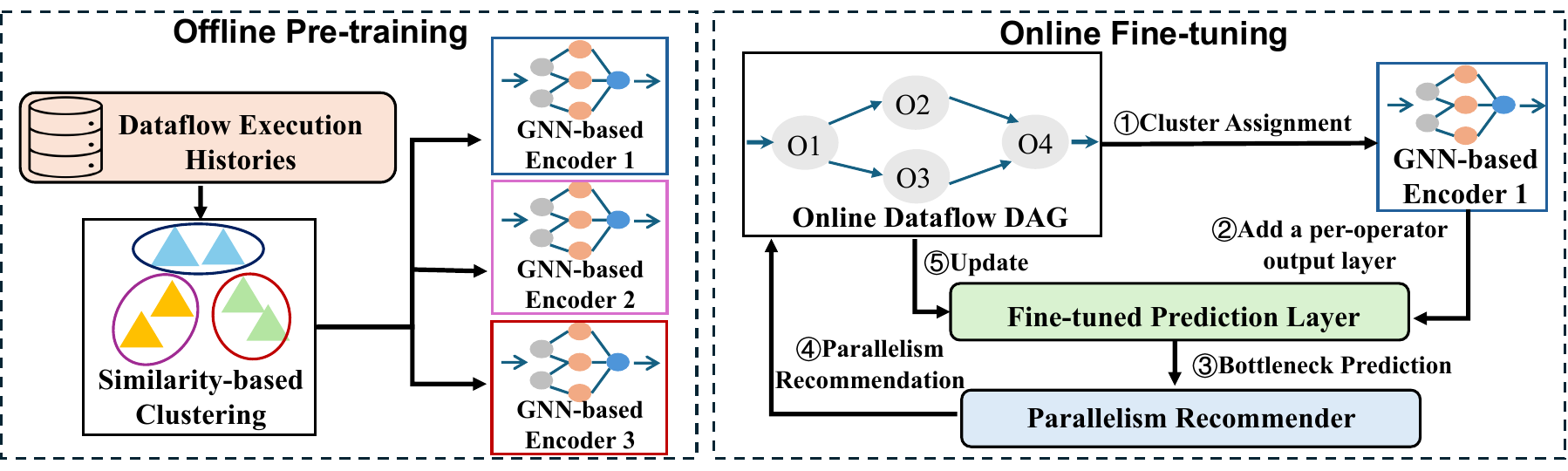}
  \vspace{-1.8em}
  \caption{Two-Phase Parallelism Tuning.}
  \vspace{-2em}
  \label{fig:overview}
\end{figure*}

\section{Overview of StreamTune}
\label{section:overview}

Figure~\ref{fig:overview} provides an overview of \sysname, which conducts parallelism tuning for streaming jobs through two phases: offline pre-training and online fine-tuning.

During the offline pre-training phase, execution histories of streaming jobs are clustered based on the structural similarity of their dataflow DAGs.
Graph Neural Networks (GNNs) are then used to encode the dataflow DAG structures, with training performed in a supervised learning framework for each cluster. 
Unlike prior methods that rely on regression tasks to train GNNs for estimating \textit{job-level} performance metrics such as latency and throughput~\cite{agnihotri2024zerotune}, our framework adopts a classification task, utilizing \textit{operator-level} bottleneck indicators as training labels.
After pre-training, the GNN-based encoders 
for the dataflow DAGs 
are trained to capture coarse correlations between dataflow structures, operator parallelisms, and bottleneck indicators. 
This pre-trained knowledge serves as a foundation for efficient online fine-tuning tailored to the target streaming job.


In the online fine-tuning phase, the target streaming job is assigned to its nearest cluster, and the corresponding pre-trained GNN-based encoder is retrieved.
A prediction layer is then added to classify the operator embeddings generated by the encoder, enabling fine-tuning for the bottleneck prediction.
Next, parallelism degrees are recommended for operators in the topological order of the dataflow DAG, guided by the bottleneck prediction.
The streaming job is redeployed with the recommended parallelisms, and its runtime feedback is collected to refine the prediction layer.
This iterative process continues until the bottleneck is fully mitigated for the streaming job or the recommended parallelism degrees
remain unchanged across iterations.

A monotonic constraint is applied to the bottleneck prediction layer during the fine-tuning phase, ensuring alignment with the observed behavior in streaming systems, where an operator's processing ability monotonically improves as its parallelism degree increases.
Note, only the prediction layer is updated during fine-tuning, while the pre-trained GNN-based encoder remains unchanged.

\noindent \uline{\textbf{Strategy for Handling Operator Parallelism.}}
\sysname employs a tailored strategy to handle operator parallelism to facilitate the tunning process.
In the pre-training phase, parallelism degrees recorded in historical data are treated separately from other operator features. 
Parallelism is incorporated into the model only after all other features are encoded during GNN training, enabling the generation of operator embeddings decoupled from parallelism.
These parallelism-agnostic embeddings are then used to guide the recommendation of operator parallelism in the fine-tuning phase.

\revise{
\noindent \uline{\textbf{Resource Considerations.}}
\sysname prioritizes CPU as a critical resource in parallelism tuning over network I/O and storage, as it is generally recognized as the primary bottleneck in numerous streaming applications. These applications, which often involve computationally intensive operations like aggregations, windowing, and stateful processing, place significant demands on CPU resources. 
As a result, the costs associated with CPU usage frequently exceed those of network I/O.
Besides, computational resources (e.g., CPU) incur a greater expenditure than storage resources (e.g., HDD, SSD) in modern cloud computing environments, highlighting the importance of CPU optimization.
}

\section{methodology}

In this section, we detail the technical design of the pre-training and fine-tuning framework employed in \sysname.
Following this, we describe our clustering approach for dataflow DAGs, which uses Graph Edit Distance to group structurally similar DAGs effectively.

\subsection{Pre-train GNN-based Encoders}
In the pre-training phase, dataflow DAGs of streaming jobs are modeled using Graph Neural Networks (GNNs) to capture their graph-structured characteristics.
We first introduce the basic concepts of GNNs and then describe the pre-training process, where a GNN-based encoder is trained to predict a binary indicator for operator-level bottleneck identification.

A GNN~\cite{DBLP:conf/iclr/XuHLJ19,DBLP:conf/nips/HamiltonYL17,xu2018representation} is a deep learning model to process graph-structured data by learning node representations through iterative messages passing~\cite{gilmer2017neural} between neighboring nodes. 
GNNs effectively model complex relationships and dependencies within a graph, making them suitable for tasks such as node classification~\cite{zhou2019meta, abu2020n}, link prediction~\cite{pareja2020evolvegcn, zhang2018link}, and graph classification~\cite{pareja2020evolvegcn, allamanis2018learning, huang2019text}.
Formally, let $G=(V, E)$ be a graph, where $V$ is the set of nodes and $E$ is the set of edges. Each node $v \in V$ has an initial feature vector $\mathbf{h}^{(0)}_{v}$. At each iteration, the message $\mathbf{h}^{(t)}_{v}$ received by node $v$ at time step $t$ is computed by aggregating the features vectors $\mathbf{h}^{(t-1)}_u$ of its neighbors $\mathcal{N}(v)$:
\begin{equation}\label{eq:aggregation}
	\mathbf{m}^{(t)}_{v} = \operatorname{AGG}(\{\mathbf{h}^{(t-1)}_u: u \in \mathcal{N}(v)\}).
\end{equation}
The node $v$ then updates its feature vectors by combining its previous feature vector $\mathbf{h}^{(t-1)}_v$ with the aggregated message $\mathbf{m}^{(t)}_{v}$:
\begin{equation}\label{eq:update}
	\mathbf{h}^{(t)}_v = \operatorname{UPDATE}(\mathbf{h}^{(t-1)}_v, \mathbf{m}^{(t)}_{v}).
\end{equation}
After $T$ iterations, the final node representation $\mathbf{h}^{(T)}_v$ integrates information from both the node and its surrounding neighbors, allowing the model to effectively capture the structural properties of the graph and the relationships among nodes.



\begin{table}[tb]
\centering
\caption{Static Features of Dataflow Operators.}
\scalebox{0.8}{
\begin{tabular}{@{}lll@{}}
\toprule
Name & Type & Description \\ \midrule
Operator Type & Categorical & Type of operator (eg., FlatMap, Filter)\\
Window Type & Categorical & Shifting strategy (tumbling/sliding) \\
Window Policy & Categorical & Windowing strategy (count/time) \\
Window Length & Numeric & Size of the window \\
Sliding Length & Numeric & Size of the sliding interval \\
Join Key Class & Categorical & Join key data type \\
Aggregate Class & Categorical & Aggregation data type \\
Aggregate Key Class & Categorical & Aggregation key data type \\
Aggregate Function & Categorical & Aggregation function (eg., min, avg) \\
Tuple Width In & Numeric & Input tuple width \\
Tuple Width Out & Numeric & Output tuple width \\
Tuple Data Type & Categorical & Type of tuple \\\bottomrule
\end{tabular}
}
\label{tbl:feature}
\end{table}

\noindent \uline{\textbf{GNN Usage for Operator-level Prediction.}}
In a dataflow DAG, nodes represent streaming operators (e.g., FlatMap or Filter), and directed edges capture computational dependencies and data flow between these operators.
\sysname encodes this structure using a GNN, which iteratively propagates and aggregates information along edges.
This approach allows each operator's representation to encode both its local features and the contextual information from its upstream and downstream operators.
The resulting node embeddings effectively capture operator-specific characteristics and the structural context of the entire DAG. 
These embeddings are then passed through a two-layer Multilayer Perceptron (MLP) with a sigmoid function to predict the operator-level bottleneck indicators.

Note, this approach differs from ZeroTune~\cite{agnihotri2024zerotune}, which combines all operator embeddings into a single summary vector to represent the overall features of the dataflow.
The summary vector is then fed into a regression model to predict job-level metrics.
By focusing on dataflow-level aggregation, ZeroTune overlooks fine-grained operator-level details, which may lead to suboptimal parallelism recommendations for individual operators.

\noindent \uline{\textbf{Operator-level Bottleneck Identification.}}
To facilitate the pre-training phase, we introduce a strategy \revise{(as outlined in Algorithm~\ref{alg:bottleneck})} for systematically analyzing and labeling operator-level bottleneck indicators in historical streaming jobs.
These indicators identify operators whose processing abilities are insufficient, which leads to dataflow-level backpressure.
An operator is labeled as $1$ if it is determined to be a bottleneck and $0$ otherwise.
The 
labeling process is described as follows:

\revise{
1) Initially, all operators are assigned a default label of -1, indicating unlabeled status (Line~1).
}

2) If no backpressure is observed at the job level, all operators are labeled as $0$, indicating no bottleneck operators~\revise{(Line~2-6)}.

3) If backpressure occurs at the job level, the \textit{cascading effect} must be considered. 
Operators under backpressure that have no downstream operators experiencing backpressure are identified first~\revise{(Line~7)}.
For these operators, the resource utilization of their downstream operators is examined \revise{(Line 8-16).
If the resource utilization $\mathcal{R}(d)$ of a downstream operator $d$ exceeds the threshold $T$ (e.g., CPU load exceeding 60\%), they are labeled as bottleneck ($L(d) \gets 1$); otherwise, they are labeled as non-bottleneck ($L(d) \gets 0$)}.
Other operators within the same dataflow DAG remain unlabeled, as the presence of job-level backpressure alters their upstream data rates, making it inconclusive to determine whether their processing ability is sufficient given their current parallelism degrees.

\begin{figure}[tb]
  \centering
  \vspace{-1.5em}
  \includegraphics[width=0.78\linewidth]{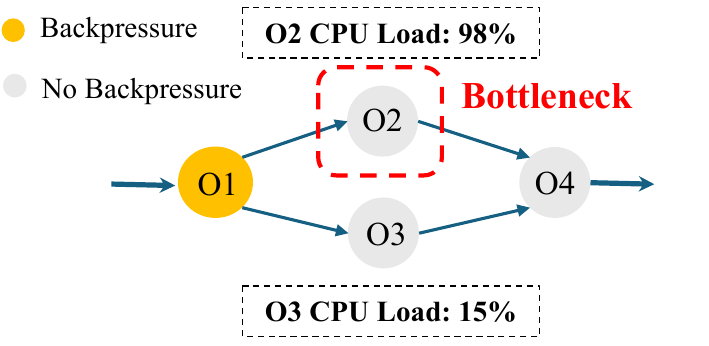}
  \vspace{-0.8em}
  \caption{\revise{Example of Bottleneck Identification.}}
  \label{fig:bottleneck}
  \vspace{-2.5em}
\end{figure}

Fig.~\ref{fig:bottleneck} provides an example of bottleneck identification in a dataflow DAG. In this DAG,  operator O1 is under backpressure due to the limited processing ability of its downstream operators, O2 and O3.
The CPU utilization of O3 is measured at $15\%$, whereas O2 exhibits a significantly higher CPU load of $98\%$, exceeding the predefined threshold.
\revise{As a result, O2 is identified as the bottleneck and labeled as  $1$; O3 and O4 are labeled as 0.}

\begin{algorithm}[t]
  \caption{\revise{Bottleneck Identification.}}
  \label{alg:bottleneck}

   \setlength{\abovecaptionskip}{0.cm}
  \SetAlgoLined
  \KwIn{
  A dataflow DAG $g=(O, E)$,
  Resource Metric $\mathcal{R}$,
  Threshold T
  }
  \KwOut{Bottleneck Labels $L(o)$ for all $o \in O$}
  
  Initialize $L(o) \gets -1$ for all $o \in O$\;

  \If{no backpressure observed}{
        \For{each operator $o \in O$} {
            $L(o) \gets 0$\;
        }
    }

    $O_b \gets$ operators $\in O$ under backpressure and no downstream operators experiencing backpressure\;

    \For{each operator $o \in O_b$ } {
        \For{operator $d \in D(o)$ as $o$'s downstream operators} {
        \eIf{$\mathcal{R}(d) > T$ }{
        $L(d) \gets 1$\;
        }
        {
        $L(d) \gets 0$\;
        }
        }
        
    }

\end{algorithm}

\noindent \uline{\textbf{Initial Feature Vector Construction.}}
To construct the initial feature vector $\mathbf{h}_v^{(0)}$,  operator features are categorized into two types: dynamic and static.
Dynamic features capture time-varying characteristics of the operator during dataflow execution. 
In our case, two dynamic features are considered:
source rates, which are controlled by external data sources, and operator parallelism, which is adjusted during tuning.
Note, only the \revise{\textit{first-level downstream operators}} have non-zero source rates.
In contrast, static features represent fixed, context-independent properties of the operator that remain unchanged during execution.
These features capture the inherent characteristics of the operator and are typically transferable~\cite{hilprechtonemodel, wu2021unified, agnihotri2024zerotune}, enabling the transfer of parallelism tuning knowledge from prior tasks to new workloads.
A list of static features for dataflow operators is provided in Table~\ref{tbl:feature}.

We construct the initial feature vector $\mathbf{h}_v^{(0)}$ for a dataflow operator $v$ using all its static features and one dynamic feature, the source rate.
Operator parallelism is excluded at this stage, as it is adaptively adjusted during the tuning process for a target streaming job.
For categorical features listed in Table~\ref{tbl:feature}, we apply one-hot encoding to represent each feature as a vector.
For numerical features, we normalize them to a continuous space $[0, 1]$ using min-max uniform scaling~\cite{LlamaTune}.
The encoded static features are then concatenated to form the initial feature vector $\mathbf{h}^{(0)}_{v}$.

\noindent \uline{\textbf{Parallelism-aware Update Strategy.}}
To incorporate the parallelism degree $\mathbf{p}_v$ as a dynamic feature for a dataflow operator $v$ at each time step $t$, the node representation is updated by combining the original feature vector $\mathbf{h}^{(t)}_v$ computed from Eq.~\ref{eq:update} as follows:
\begin{equation}
	\mathbf{h}'^{(t)}_v = \operatorname{FUSE}(\mathbf{h}^{(t)}_v || \mathbf{p}_v).
\end{equation}
The function $\operatorname{FUSE}(*)$  performs two key operations:

1) it integrates the feature to be adjusted (i.e., parallelism) with other operator features to enhance
bottleneck prediction.

2) it applies a non-linear transformation that ensures the dimensionality of the updated node representation $\mathbf{h}'^{(t)}_v$ remains unchanged after integrating parallelism.
This consistency enables $\mathbf{h}'^{(t)}_v$ to seamlessly participate in subsequent message-passing iterations.

Here, $\mathbf{h}^{(t)}_v$ and $\mathbf{h}'^{(t)}_v$ at the message passing iteration $t$ are referred to as  the \textit{parallelism-agnostic} and \textit{parallelism-aware} feature vector, respectively.

\begin{algorithm}[t]
  \caption{Online Parallelism Tuning.}
  \label{alg:tune}

   \setlength{\abovecaptionskip}{0.cm}
  \SetAlgoLined
  \KwIn{
  A streaming task with its dataflow DAG $g$, 
  The physical maximum parallelism $p_{max}$,
  A set of cluster centers $SC = $ $\{sc_1, sc_2, ..., sc_K\}$ and their corresponding GNN-based Encoders $GE = \{enc_1, enc_2,..., enc_K\}$
  }
	
	Assign the target DAG $g$ to its nearest cluster:
	$$c=\arg \min\limits_{k\in \{1,2,...K\}}\revise{\mathbf{dist}}(g, sc_k);$$ \\
	
	Retrieve the corresponding GNN Encoder $enc_c$;\\
	
	$\mathcal{T}$ = ConstructWarmUpDataset($enc_c$)\;
%
%


  \Do{no backpressure \textnormal{\textbf{and}} recommended parallelisms differ from current ones}{
	Fit a monotonic model $\mathcal{M}_f$ to $\mathcal{T}$;\\
    \For{each operator $v$ in $G$ in $g$'s topological order}
    {
      Retrieve its \textit{parallel-agnostic} feature vector $\mathbf{h}_v$ via $enc_{c}(G)$; \\
      Set parallelism for $v$ as: 
      $p^{rec}_{v} \gets  \min \{ p \le p_{max} \mid \mathcal{M}_{f}(\mathbf{h}_v, p) = 0 \}$\;
    }
    
    $\Delta\mathcal{T}$ = ReDeployAndMonitorBackpressure($\{p^{rec}_{v}\}$)\;

    $\mathcal{T} \gets \mathcal{T} \cup \Delta\mathcal{T}$ \;
 
 
    
  }

\end{algorithm}

\noindent \uline{\textbf{Loss Function.}} 
To pre-train the GNN-based encoders with bottleneck labels, we utilize the binary cross-entropy loss function~\cite{mannor2005bceloss}, denoted as $\mathcal{L}_v$, for each labeled dataflow operator $v$.
This loss function measures the difference between the predicted probability distribution of the bottleneck indicator and the true label distribution.
For a given set of operators $\mathcal{O}$ in the pre-trained dataset, 
we denote the labeled set of operators as $\mathcal{O}_{label}$.
The total loss for the labeled set is computed as the average loss across all labeled operators: $\mathcal{L}_{total} = \frac{1}{|\mathcal{O}_{label}|} \sum\nolimits_{v\in \mathcal{O}_{label}}{\mathcal{L}_v}$.

\subsection{Online Fine-tuning}\label{sec:online}

We begin by presenting the workflow of the online fine-tuning phase and then provide a detailed discussion of its key component, specifically the fine-tuned model for the bottleneck prediction with a monotonic constraint, including its formal definition and the rationale behind the model selection.

This phase employs an iterative process for parallelism tuning of a target streaming task's dataflow DAG $g$, as outlined in Algorithm~\ref{alg:tune}.
First, we assign the dataflow DAG $g$ of the job to the nearest cluster $c$, using a distance measure \revise{$\mathbf{dist}$} to quantify the similarity between $g$ and the cluster centers (Line~1).
The corresponding GNN encoder $enc_c$ for the cluster $c$ is then retrieved to encode $g$ (Line~2).
Prior to initiating the iterative tuning process, a warm-up training dataset for the fine-tuned model is prepared~(Line~3).
This involves sampling a subset of dataflow DAGs from cluster $c$ and using $enc_c$ to generate embeddings for the dataflow operators from the sample data. These embeddings, along with their corresponding bottleneck labels, form the training dataset $\mathcal{T}$ for fine-tuning (Line~3).

The tuning proceeds by fitting a fine-tuning model $\mathcal{M}_f$ with monotonic constraint to the dataset $\mathcal{T}$ (Line~5) and recommending parallelism degrees for each dataflow operator based on $g$'s topological order (Line~6).
For each operator $v$, we compute its \textit{parallelism-agnostic} feature vector $\mathbf{h}_v$ using $enc_c$ (Line~7).
At this stage, the \textit{parallelism-agnostic} feature vector is used instead of the \textit{parallelism-aware} one, as $v$'s parallelism has not yet been recommended.
Then, the minimum parallelism degree that prevents the operator from becoming a bottleneck is identified through a search based on $\mathcal{M}_f$'s predictions (Line~8).
Leveraging the monotonic constraint of $\mathcal{M}_f$, this search can be implemented as a binary search.
Once parallelism recommendations for all operators are generated, the streaming job is redeployed with these parallelisms, and corresponding bottleneck labels are collected (Line~10).
The dataset $\mathcal{T}$ is then updated with these new feedbacks to further refine $\mathcal{M}_f$ in the next iteration (Line~11).
The iterative process continues until either no job-level backpressure is observed for the streaming job $g$ or the recommended parallelisms no longer differ from the previous iteration.

\begin{figure}[tb]
  \centering
  \includegraphics[width=0.8\linewidth]{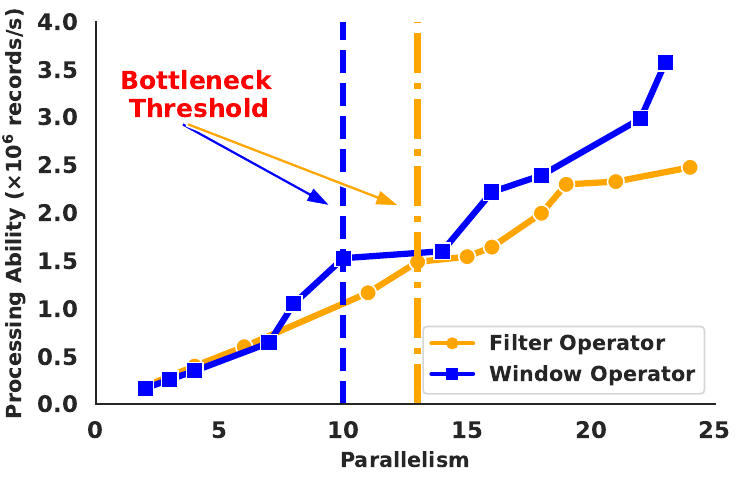}
  \vspace{-1.5em}
  \caption{\small Relationship Between Parallelism and Processing Ability.}
  \vspace{-2em}
  \label{fig:monotonic}
  
\end{figure}

\noindent\uline{\textbf{Monotonic Constraint of $\mathcal{M}_f$.}}
%
The monotonic constraint of $\mathcal{M}_f$ is based on the assumption that the probability of a dataflow operator becoming a bottleneck monotonically decreases as its parallelism degree increases.
This assumption reflects the observed behavior in stream processing systems, where increasing an operator's parallelism degree enhances its processing ability.

To validate this, we analyze a typical streaming job  from~\cite{agnihotri2024zerotune} executed on Apache Flink.
This job consists of two operators: a filter operator that processes streams based on specific conditions and a subsequent window operator that aggregates data.
To examine the relationship between parallelism and processing ability for each operator, we fix the source rate and the parallelism degree of one operator while varying the parallelism degree of the other. 
The observed relationships are depicted in Fig.~\ref{fig:monotonic}.
As shown, the processing ability of both operators increases with their parallelism degrees.
When their parallelism degrees are below specific thresholds ($14$ for the filter operator and $10$ for the window operator), the operators cause backpressure and become bottlenecks. 
Beyond these thresholds, the operators can handle incoming data without bottlenecking. 
These results indicate the presence of a bottleneck threshold within the parallelism value domain. Below this threshold, operators are prone to backpressure, while above it, they can process data efficiently.



We now formally define the monotonic constraint of $\mathcal{M}_f$.
For an input $\mathbf{x}=[\mathbf{h}, p]$ where $\mathbf{h}$ represents as the \textit{parallel-agnostic} vector, and $p$ is the recommended parallelism, 
$\mathcal{M}_f$ could be seen as a classification model that estimates the probability $p(\mathbf{x})$ that the instance $\mathbf{x}$ belongs to class $1$.
This probability is defined as $p(\mathbf{x}) = \mathbb{P}(y=1| \mathbf{x})$, where $y \in \{0,1\}$ is the class label.
Class 1 indicates the operator is a bottleneck causing backpressure, while class 0 indicates it is not.
The monotonic constraint ensures that $p(\mathbf{x})$ is non-increasing with respect to the parallelism $p$. 
Formally, for any two inputs ($\mathbf{h}^i$, $p^i$) and ($\mathbf{h}^j$, $p^j$) such that $p^i \le p^j$, the following condition holds:
$p(\mathbf{h}^i, p^i) \ge p(\mathbf{h}^j, p^j)$. This constraint captures the observed system behavior that increasing parallelism reduces the likelihood of an operator becoming a bottleneck.

\noindent\uline{\textbf{Model Choice For $\mathcal{M}_f$.}}
The model choice of $\mathcal{M}_f$ must achieve a balance between accuracy and computational efficiency, given the need for low overhead in the online tuning process. Considering the difficulty neural networks encounter in enforcing monotonicity~\cite{DBLP:conf/nips/LiuHZ020,DBLP:conf/icml/RunjeS23}, we propose two lightweight models as candidates for constructing $\mathcal{M}_f$:


  \textbf{(a) SVM}~\cite{cortes1995svm}: Given $\mathbf{x}=[\mathbf{h}, p]$ as a data point, the Support Vector Machine (SVM) seeks to find a hyperplane that separates data points from different classes with the maximum margin using a decision function. 
To effectively capture nonlinear relationships in $\mathbf{h}$, we apply the kernel trick~\cite{boser1992kerneltrick}, transforming $\mathbf{h}$ into a higher-dimensional feature space.
The decision function is expressed as:
\begin{equation}
	f(\mathbf{x})= \mathbf{w}_e^T\phi(\mathbf{h}) + w_{p}p+b,
\end{equation}
where $\phi(\mathbf{h})$ is the feature mapping induced by the chosen kernel function (e.g., RBF, polynomial), $\mathbf{w}_e$ is the weight vector in the transformed feature space, $w_p$ is the weight associated with $p$ and $b$ is the bias term.
To account for regularization and misclassification, its objective function incorporates a regularization coefficient $C$ and slack variables $\xi_i$.
To enforce the monotonicity constraint on $p$, the objective function is reformulated as follows:
\begin{equation}
	\begin{aligned}
    &\min_{\mathbf{w}_e, w_p, b, \boldsymbol{\xi}} \quad 
    \frac{1}{2} \|\mathbf{w}_e\|^2 + \frac{1}{2} w_p^2 + C \sum_{i=1}^n \xi_i \\
    &\text{subject to} \quad 
    y_i \left( \mathbf{w}_e^\top \phi(e_i) + w_p p_i + b \right) \geq 1 - \xi_i, \\
    &\hspace{4.5em} \xi_i \geq 0, \quad w_p \leq 0.
\end{aligned}
\end{equation}
Here, the constraint $w_p \le 0$ enforces the monotonicity constraint, ensuring that as the parallelism $p$ increases, the probability of the operator being a bottleneck decreases.


\textbf{(b) XGBoost}~\cite{chen2016xgboost}: XGBoost is a scalable machine learning model that builds an ensemble of weak decision trees through gradient boosting~\cite{ke2017lightgbm}.
However, standard decision trees do not naturally enforce monotonic relationships between features and the target variable.
To address this, we modify the tree-building process to ensure that the resulting ensemble adheres to the monotonic constraint. This involves adjustments to both tree splitting and leaf value assignment.
For each potential split in a decision tree, the algorithm evaluates whether it complies with the monotonicity constraint. 
Splits that violate the constraint are penalized by setting their gain to $-\infty$, effectively excluding them from consideration. 
For leaf value assignment, the monotonic constraint ensures that for any two leaf nodes, if one corresponds to a higher value of the constrained feature (i.e., parallelism $p$), the predicted value associated with that leaf must respect the monotonic order relative to the other. 
This approach ensures that the final ensemble model aligns with the monotonic behavior required for accurate and interpretable predictions.

\subsection{Clustering on Dataflow DAGs}
\noindent \uline{\textbf{Dataflow Distance Measure.}} 
%
To effectively cluster extensive historical dataflow datasets, it is essential to define a distance measure between dataflow DAGs that ensures the validity of the clustering process. 
This measure must satisfy the \textit{triangle inequality} property to avoid paradoxical results. The triangle inequality guarantees that the direct distance between two objects is always less than or equal to the sum of the distances through an intermediate object.
Formally, for any three dataflow DAGs, $g_1$, $g_2$, and $g_3$, the distance measure \revise{$\mathbf{dist}$} must satisfy the following condition: 
\begin{equation} 
\revise{\mathbf{dist}}(g_1, g_3) \leq \revise{\mathbf{dist}}(g_1, g_2) + \revise{\mathbf{dist}}(g_2, g_3).
\end{equation}

To this end, we employ Graph Edit Distance (GED)~\cite{gouda2016csi_ged,sanfeliu1983distance,kim2021boosting} as it satisfies the triangle inequality and provides an interpretable similarity measure between graphs.
Although the computation of GED is NP-hard\cite{zeng2009comparing,ma2002computing}, it remains practical for our scenario due to the relatively small scale of dataflow DAGs, which typically contain fewer than 20 nodes and edges.
Formally, the GED between two DAGs $g_1$ and $g_2$, denoted as $\operatorname{ged}(g_1, g_2)$, is defined as the minimum number of \textit{edit operations} required to transform $g_1$ into $g_2$.
While GED is commonly used for undirected graphs, it can be extended to directed graphs by incorporating additional edit operations tailored to dataflow DAGs. 
In addition to standard operations such as node insertion, node deletion, edge insertion, and edge deletion, we introduce two new operations:
\begin{itemize}[left=0.5em, topsep=0pt]
\item 
\textit{Operator Type Modification}: Change the operator type of an existing node (e.g., from filter to join).
\item
\textit{Edge Direction Modification}: Reverse the direction of an existing edge.
\end{itemize}

\noindent \uline{\textbf{Clustering with GED.}} 
With the distance metric defined, the K-means clustering algorithm~\cite{ahmed2020k} is applied to group similar dataflow DAGs, treating each DAG as a data point in the clustering space.
This clustering algorithm partitions the dataset into $k$ clusters, assigning each DAG to the cluster whose centroid is nearest.
The clustering process involves three main steps: 1) \textit{Initialization}: Randomly select $k$ DAGs from the dataset as the initial cluster centroids. 2) \textit{Assignment}: Compute the GED between each DAG and all centroids, assigning the DAG to the cluster with the closest centroid. 3) \textit{Update}: Recalculate the centroids of the clusters based on the current assignments.  
Steps 2 and 3 are repeated iteratively until the centroids converge or a predefined maximum number of iterations is reached.



A key challenge in graph clustering lies in the absence of a direct method for computing the centroid of a set of graphs during the update step. 
Ideally, the centroid should capture the most central or representative data points of the cluster. 
Unlike numerical data, graphs cannot be easily 	``averaged''.
One potential solution is the concept of the median graph~\cite{jiang2001median}, which minimizes the total GED to all other graphs in the cluster. 
However, this approach requires calculating the GED for every pair of graphs in the cluster, which can be computationally expensive.
To address this challenge, we propose the concept of an approximate median graph, referred to as the \textit{similarity center}. 
This graph is defined as the most frequently occurring dataflow DAG identified through graph similarity search~\cite{chang2020graphsimilarsearch} across all DAGs within a cluster. 
Due to its proximity to other graphs in the cluster, we argue that it provides an effective approximation of the cluster's central graph. 
Its formal definition is provided as follows:

\begin{definition}[Graph Similarity Search]
	Given a cluster of dataflow DAGs $\mathcal{G}$, a query graph $q$ and a GED distance threshold $\tau$, a graph similarity search finds all the DAGs whose GED to $q$ do not exceed $\tau$. i,e, $ {Sim}_{q,\tau} = \{g\in \mathcal{G} |  ged(q,g) \le \tau\}$. 
\end{definition}

\begin{definition}[Similarity Center]
	Given a cluster of dataflow DAGs $\mathcal{G}$ and a DAG $g \in \mathcal{G}$, the appearance count $\mathcal{C}_g$ of $g$ is defined as the number of times $g$ appears in the similarity search results for all DAGs in $\mathcal{G}$, using a GED threshold $\tau$. 
	Formally, this count is expressed as: $\mathcal{C}_g=\sum_{g' \in \mathcal{G}} \mathbb{I}(g \in {Sim}_{g', \tau})$, where $\mathbb{I}(\cdot)$ is the indicator function.
	The similarity center of the cluster $\mathcal{G}$ can be defined as:
	\begin{equation}
		\mathcal{G}_{sc} = \arg \max\limits_{g \in \mathcal{G}} \mathcal{C}_g.
	\end{equation} 
\end{definition}

To efficiently compute the similarity center of a DAG cluster, it is crucial to accelerate the graph similarity search.
While graph similarity search has been extensively studied~\cite{zhao2013partition, kim2019inves, zheng2014efficient, chang2020graphsimilarsearch}, selecting the most appropriate approach for our setting remains challenging.
A common strategy employs a \textit{filtering-and-verification} approach to reduce the computational burden of GED calculations~\cite{zhao2013partition, kim2019inves}. 
This strategy builds an index over the graph dataset to prune irrelevant graphs during the filtering phase, followed by a verification phase to identify the qualifying graphs.
In \sysname, we employ the AStar$^+$-LSa algorithm~\cite{chang2020graphsimilarsearch} for graph similarity search due to two key advantages:

\noindent 1) \textbf{Index-Free Operation:} Unlike index-based methods, AStar$^+$-LSa avoids reliance on pre-build indices, which can become outdated as DAGs within clusters evolve across the update iterations, incurring significant rebuilding overhead.
Instead, AStar$^+$-LSa optimizes the graph similarity search by directly refining the GED computation. 

\noindent 2) \textbf{Efficiency in GED Computation}: 
AStar$^+$-LSa optimizes GED computation directly using a best-first search strategy. 
The algorithm begins with an initial node mapping~\cite{abu2015exact}, 
building correspondences between nodes of the query and target graphs 
to compute GED through minimal edit operations. 
Partial mappings, representing incomplete correspondences for subsets of nodes, are incrementally explored during the search process. 
For each partial mapping, a tight lower bound is calculated by considering node and edge similarities. Branches with bounds exceeding the threshold are pruned, effectively reducing the search space and memory usage.
The best-first search strategy not only accelerates similarity search but also improves the speed of GED computation itself.
This dual efficiency makes it highly effective for \sysname's clustering of dataflow DAGs, substantially accelerating all three steps of the K-means clustering process.

\section{Experiments}
In this section, we present the evaluation results of \sysname, assessing its effectiveness and efficiency in performing automatic parallelism tuning for stream processing systems. 
The evaluation focuses on addressing the following questions:

\begin{itemize} [leftmargin=*]
\item  \textbf{\revise{Exp-Q1}:} Can \sysname recommend optimal parallelism for streaming jobs to handle varying source rates while conserving CPU resources?  (See Sec.~\ref{sec: exp_paralleism}).

 \item \textbf{\revise{Exp-Q2}:} Does \sysname minimize the number of reconfiguration operations required during the tuning process\revise{, and how long does \sysname take to adapt to an unseen workload?} (See Sec.~\ref{sec: exp_reconf})

 \item \textbf{\revise{Exp-Q3}:} Can \sysname effectively eliminate bottlenecks compared to existing methods? (See Sec.~\ref{sec: exp-bottleneck})

\item \textbf{\revise{Exp-Q4}:} Can \sysname demonstrate generality by performing effectively across different stream processing systems?  (See Sec.~\ref{sec: exp_generality})

 \item \textbf{\revise{Exp-Q5}:} Which model performs best in the fine-tuning phase of \sysname? (See Sec.~\ref{sec: exp_abalation})

 \item \revise{\textbf{Exp-Q6}: What is the computational cost of \sysname in both online tuning and offline pre-training phases? (See Sec.~\ref{sec: exp_overhead})}

\item \revise{\textbf{Exp-Q7}: How does StreamTune impact CPU utilization dynamics during the tuning process? (See Sec.~\ref{sec:cpu_utilization}) } 

\item  \textbf{\revise{Exp-Q8}}: Does clustering historical dataflow DAGs improve tuning efficiency, and how does graph similarity search enhance the clustering process? (See Sec.~\ref{sec: exp_abalation})

\end{itemize}

\vspace{-0.5em}
\subsection{Experimental Setup}

\noindent \textbf{Environment.} 
All experiments were conducted on Debian 11 with Linux kernel version 5.56.
Flink experiments were conducted on up to two physical machines, each with $80$ Intel(R) Xeon(R) Gold 6230 CPUs running at 2.10GHz and $380$GB of RAM. 
We use Apache Flink 1.16 configured with $50$ \revise{TaskManagers}, each with $2$ slots (maximum parallelism per operator=$100$).
\revise{Please note each task slot represents a fixed subset of resources of Flink's TaskManager.}
Timely experiments were conducted on a single physical machine with $128$ Intel(R) Xeon(R) Platinum 8336C CPUs running at 2.30GHz and $1.5$TB of RAM, \revise{worker threads were evenly distributed across CPU cores to prevent potential contention.}
We use Timely Dataflow v0.10.0 compiled with Rust $1.82.0$ and configured with ten workers.

\begin{table}[]
\caption{Source Rate Units of Different Streaming Jobs}
\vspace{-0.5em}
\scalebox{0.69}{
\begin{tabular}{@{}lllllllll@{}}
\toprule
& \multicolumn{2}{c}{Bids} & \multicolumn{2}{c}{Auctions} & \multicolumn{2}{c}{Persons} & \multicolumn{2}{c}{PQP Source} \\
& \multicolumn{1}{c}{Flink} & \multicolumn{1}{c}{Timely} & \multicolumn{1}{c}{Flink} & \multicolumn{1}{c}{Timely} & \multicolumn{1}{c}{Flink} & \multicolumn{1}{c}{Timely} & \multicolumn{2}{c}{Flink} \\ 
\cmidrule(l){2-9} 
\revise{(Nexmark)}Q1 & 700K & 9M & / & / & / & / & \multicolumn{2}{l}{/} \\
\revise{(Nexmark)}Q2 & 900K & 9M & / & / & / & / & \multicolumn{2}{l}{/} \\
\revise{(Nexmark)}Q3 & / & / & 200K & 5M & 40K & 5M & \multicolumn{2}{l}{/} \\
\revise{(Nexmark)}Q5 & 80K & 10M & / & / & / & / & \multicolumn{2}{l}{/} \\
\revise{(Nexmark)}Q8 & / & / & 100K & 4M & 60K & 4M & \multicolumn{2}{l}{/} \\
\revise{(PQP)}Linear & / & / & / & / & / & / & \multicolumn{2}{l}{5K} \\
\revise{(PQP)}2-way-join & / & / & / & / & / & / & \multicolumn{2}{l}{0.5K} \\
\revise{(PQP)}3-way-join & / & / & / & / & / & / & \multicolumn{2}{l}{0.25K} \\
\bottomrule
\end{tabular}
}
\label{tab: nexmark_source_rate}
\vspace{-1.5em}
\end{table}

\noindent \textbf{Workloads.} We evaluate the performance of parallelism tuning methods using two representative streaming workloads:

\begin{enumerate} [leftmargin=*]
\item \textbf{\textit{Nexmark Benchmark}}~\cite{nexmark2024}: 
\revise{For this evaluation, we employ Nexmark queries Q1, Q2, Q3, Q5, and Q8, which represent a diverse set of streaming operators commonly encountered in real-world data processing scenarios.
Specifically, Q1 and Q2 involve stateless operators, i.e., map and filter, respectively.
Q3 introduces a stateful, record-at-a-time two-input operator, i.e., an incremental join.
Q5 and Q8 contain two distinct window operators, i.e., sliding window join and tumbling window join, respectively.}
  
\item \textbf{\textit{PQP Queries}}~\cite{agnihotri2024zerotune}:
It contains synthetic queries designed to assess the generalization and accuracy of parallelism tuning methods across diverse streaming scenarios. 
It includes linear queries, multi-way joins, and chained filters, reflecting various operator dependencies. 
The workload also features common streaming operators such as source, filter, join, and aggregate, with configurations for tumbling and sliding windows.
For this evaluation, three query templates are used: Linear (8 queries), 2-way-join (16 queries), and 3-way-join (32 queries).
\end{enumerate}

\noindent \textbf{Source Rate Simulation.}
To simulate dynamic workloads characteristic of real-world streaming jobs, 
we vary the source rate over time for the evaluated workloads using a periodic pattern.
Specifically, a basic cycle of ten 
source rates can be defined as [$3W_u$, $7W_u$, $4W_u$, $2W_u$, $1W_u$, $10W_u$, $8W_u$, $5W_u$, $6W_u$, $9W_u$], where $W_u$ denotes the source rate unit, measured in records per second.
The values of $W_u$ for different streaming jobs are provided in Table~\ref{tab: nexmark_source_rate}.
To introduce periodicity, the ten source rates are replicated, forming a sequence of $20$ source rates.
Six different permutations of this sequence are generated for each query, resulting in a total of 120 ($20 \times 6$) source rate changes per query.

\begin{figure}[t]
  \centering
  \includegraphics[width=0.9\linewidth]{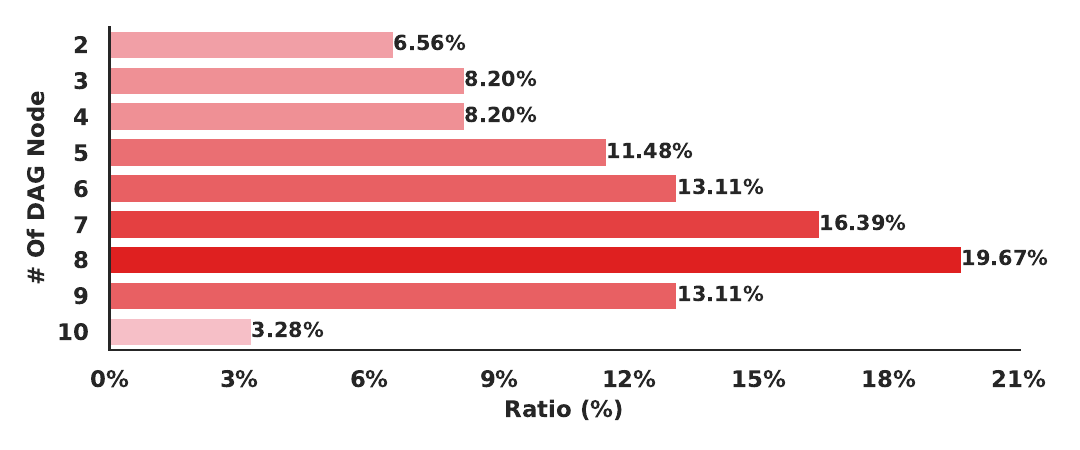}
  \vspace{-1.3em}
  \caption{\revise{Distribution of Pre-trained Dataflow DAGs}}
  \vspace{-2em}
  \label{fig: Query_Distribution}
\end{figure}

\noindent \revise{\textbf{Pre-training Setup.}
To construct the training dataset for pre-training, we utilized the execution histories of Nexmark and PQP queries. 
Fig.~\ref{fig: Query_Distribution} illustrates the distribution of pre-trained queries according to the number of logical dataflow operators. 
For source rates, we selected random values between ($1W_u$,$10W_u$) and ensured that the rates used in tuning differ from those in pre-training.
For parallelism degrees, we assigned random values from $[1, 60]$ for each dataflow operator across different queries.
Bottleneck labels for dataflow operators were determined using Algorithm~\ref{alg:bottleneck}.  To ensure stability, a $10$-minute wait is enforced between reconfigurations. This prevents fluctuations and allows the system to adapt new parallel degrees before further tuning adjustments.
}
%
The optimal number of clusters for the k-means algorithm, $k$, is determined using the elbow method~\cite{ketchen1996elbowmethod}.
For the graph similarity search employed in calculating clusters' similarity centers, the distance threshold $\tau$ is set to $5$.

\begin{figure*}[t]
  \centering
  \includegraphics[width=1\linewidth]{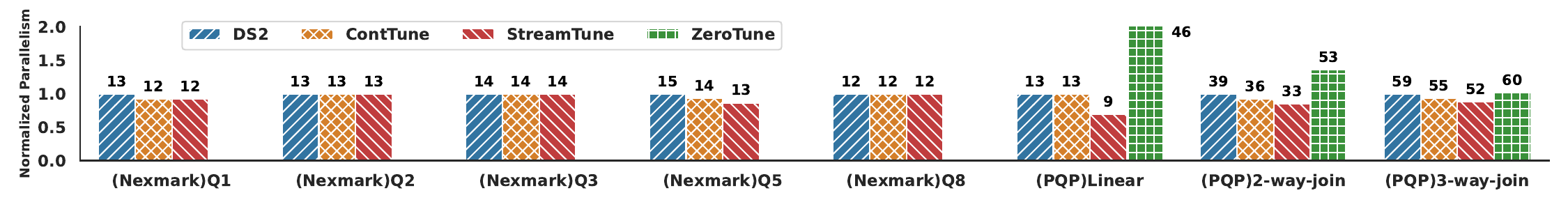}
  \vspace{-2em}
  \caption{Final parallelism recommendations by different methods when the source data rate is $10 \times W_u$ in Flink.}
  \label{fig:parallelism_number}
  \vspace{-2em}
\end{figure*}

\noindent \revise{\textbf{Reconfiguration Mechanism.}}
\revise{Our evaluation follows the DS2 approach, where the system stops and restarts for reconfiguration. The current source logic is part of the dataflow construction, allowing sources to stop naturally during reconfiguration. This enables a fair comparison of different parallelism tuning methods under controlled conditions.}

\noindent \textbf{Competitors.} We compare \sysname with three baselines:

\begin{enumerate} [leftmargin=*]
    \item \textbf{\textit{DS2}}~\cite{kalavri2018three}: A representative method based on the linearity assumption between operator parallelism and processing ability. We use its original implementation~\cite{ds2implementation}.
    \item \textbf{\textit{ContTune}}~\cite{lian2023conttune}: A Bayesian Optimization-based method that tunes the parallelism of a streaming job using its own tuning history.  We implement ContTune with scikit-learn v1.2.2 in Python 3.8. The hyperparameter $\alpha$ in its scoring function is set to $3$, following the optimal setting reported in the original experiments.
    \item \textbf{ZeroTune}~\cite{agnihotri2024zerotune}: A method that employs zero-shot learning to construct a cost model for dataflow-level performance prediction. Using this cost model, ZeroTune determines the initial parallelism degrees for a streaming job.
        Since it does not specify a parallelism tuning strategy, we sample different groups of parallelisms for dataflow operators and select the one with the lowest estimated cost.
\end{enumerate}



\subsection{Implementation}
Operator-level bottleneck detection plays a critical role in both the pre-training and fine-tuning phases. Here, we describe the implementation of this functionality in Apache Flink and Timely Dataflow.

\noindent \textbf{Apache Flink:} In Apache Flink, the operational states of dataflow operators during execution are classified into three categories: under backpressure, idle, and busy. Three built-in metrics quantify these states:
(i) \textit{backPressuredTimeMsPerSecond}, which measures the duration an operator spends under backpressure;
(ii) \textit{idleTimeMsPerSecond}, which measures the duration an operator remains idle;
and (iii) \textit{busyTimeMsPerSecond}, which measures the duration an operator is actively processing data.
A Flink operator is considered a bottleneck if its \textit{backPressuredTimeMsPerSecond} exceeds $10\%$ of the cumulative sum of these metrics over a sustained interval.

\noindent \textbf{Timely Dataflow:} 
Timely Dataflow lacks a built-in backpressure mechanism, making it unsuitable for directly detecting bottlenecks.
Instead, we define a Timely operator as a bottleneck if its input data rate falls below $85\%$ of the combined output rates of all its upstream operators.
To collect the input/output data rate of an operator,  we leverage raw log information from Timely.
However, since Timely operators are non-blocking and continuously spinning, they generate a high volume of event logs, many of which are irrelevant to our analysis. 
To address this, we modified Timely's log recorder to capture only useful events from each worker.
Specifically, we use \textit{MessagesEvent} to record runtime metrics related to data rates of physical operators.
These metrics are periodically aggregated to compute cumulative data rates for logical operators.

%

\vspace{-0.5em}
\subsection{Comparison of Recommended Parallelism} \label{sec: exp_paralleism}
We evaluate the total operator parallelism for each streaming job after several reconfigurations by different tuning methods when the source rate changes to $10W_u$. 
This metric reflects resource consumption, such as the CPU cores required during query execution. This experiment was conducted on Apache Flink, with SVM chosen as \sysname's prediction output layer during the fine-tuning phase.
Figure~\ref{fig:parallelism_number} presents the results for total parallelism. 
Note, ZeroTune is specific to PQP queries and is therefore not evaluated on the Nexmark benchmark.

Overall, \sysname achieves the lowest recommended parallelism across various streaming jobs.
ContTune generally outperforms DS2 by employing Bayesian Optimization to incorporate the tuning history of the target query.
For PQP queries, ZeroTune consistently recommends the highest parallelism, as its optimization focuses solely on minimizing dataflow performance metrics (e.g., latency and throughput) without considering resource efficiency.

Specifically, the parallelism recommended by DS2, ContTune, and \sysname are similar for Nexmark jobs such as Q1-Q3.
This is because these jobs have simple DAG structures and involve few dataflow operators.
For more complex Nexmark jobs such as Q5 and PQP jobs, \sysname consistently recommends a lower parallelism compared to DS2 and ContTune. 
This improvement is attributed to \sysname's GNN-based encoder, which effectively captures intricate dataflow patterns and operator dependencies, leveraging global historical data for more informed recommendations.
In contrast, DS2 and ContTune rely on collecting \textit{useful time} to compute the \textit{processing abilities} of each operator for their recommendations.
\revise{However, accurately measuring \textit{useful time}, which relates to CPU utilization, is intricate in real-world dataflow executions and may impact the accuracy of parallelism recommendations.}



\begin{figure*}[t]
\centering
\hspace{-0.8cm}
\subfloat[\revise{Average number of reconfigurations per tuning in Apache Flink.}]{
\label{fig:reconfiguration}
\includegraphics[trim=0.4cm 0.4cm 0cm 0.4cm, clip, scale=0.42]{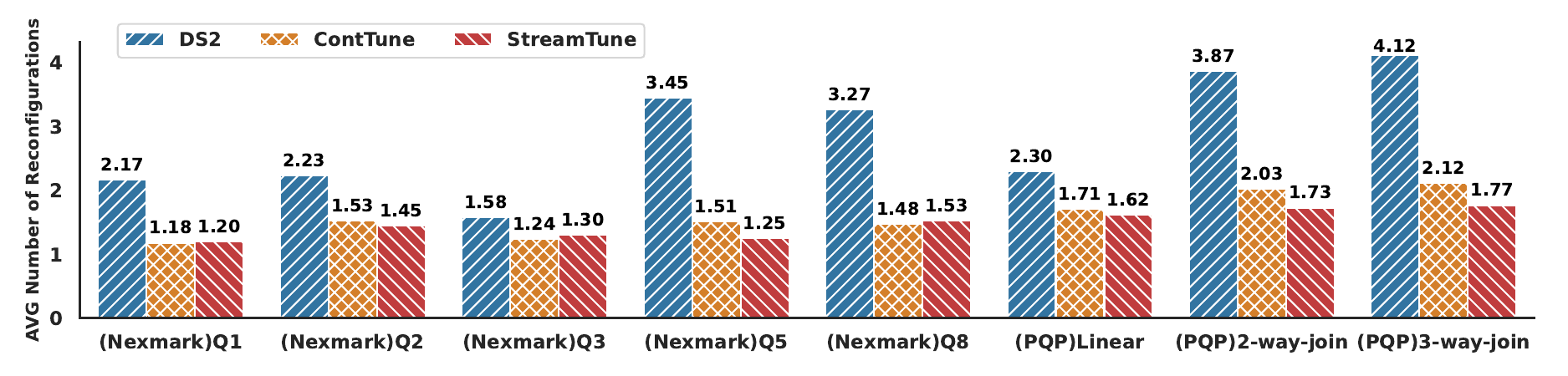}
}
\hspace{-0.6cm} 
\subfloat[\revise{\small{Case study: an unseen 2-way-join query.}}]{
\label{fig:unseen_2wayjoin}
\vspace{-3mm}
\includegraphics[trim=0.4cm 0.4cm 0cm 0.4cm, clip, scale=0.53]{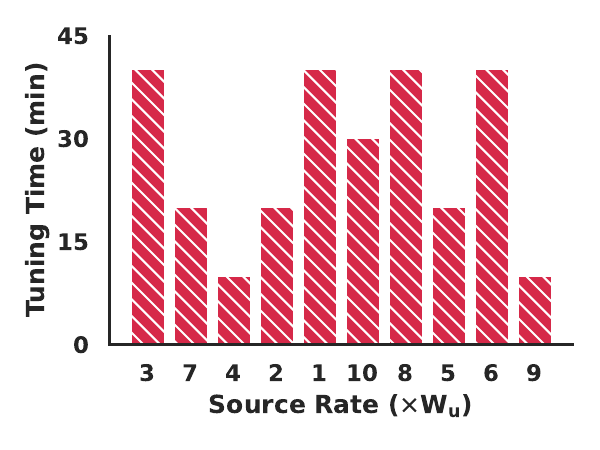}
}
\vspace{-2mm}
\caption{\revise{Reconfiguration efficiency and adaptation time of \sysname}}
\vspace{-2em}
\end{figure*}

\vspace{-0.5em}
\subsection{Comparison of Reconfiguration Frequency} \label{sec: exp_reconf}
We then report the average number of reconfigurations for each tuning process of different methods in response to changes in the source rate. 
Minimizing the number of reconfigurations is desirable in streaming processing systems, as fewer adjustments reduce fluctuations and enhance system stability.
For this evaluation, we record the total number of reconfigurations performed by each method throughout the periodic pattern on Flink and divide it by the total number of tuning processes (i.e., 120) to compute the average.
Note, ZeroTune always performs a single reconfiguration, as its cost model directly determines the initial parallelism. As a result, this evaluation focuses on DS2, ContTune, and \sysname.

Figure~\ref{fig:reconfiguration} shows the evaluation results.  
For the Nexmark benchmark, the results indicate that ContTune and \sysname require a similar number of reconfigurations, while DS2 demands significantly more.  
This discrepancy arises because ContTune and \sysname leverage historical data to guide their tuning decisions, whereas DS2 relies on a fixed formulation based on the linearity assumption to compute parallelism.
Without the benefit of historical tuning data, DS2 exhibits slower convergence.
For PQP queries, \sysname requires fewer reconfigurations compared to ContTune.
This advantage comes from ContTune's reliance on local historical data of the job to be tuned, with its focus on single-operator granularity.
Consequently, ContTune struggles with structurally complex queries. 
In contrast, \sysname leverages global historical data from similar dataflows, combined with a GNN to capture operator-level bottlenecks. 
This enables \sysname to perform more effectively for complex streaming queries, which requires fewer reconfigurations to achieve optimal parallelism.
\revise{Notably, in PQP Linear Query, \sysname can reduce the number of reconfigurations by up to 29.6\%, while simultaneously decreasing the parallelism degree by 30.8\%.}

\revise{To evaluate \sysname's adaptability to unseen workloads, we conduct a case study by isolating one 2-way-join PQP query from the pre-training phase and measuring its tuning time under periodic source rate changes over Apache Flink. 
The result in Fig.~\ref{fig:unseen_2wayjoin} shows how tuning time fluctuates across different source rates. 
The results show that StreamTune's tuning time ranges from approximately 10 to 40 minutes, with an average tuning time of around 27 minutes.
Note, the reported tuning time includes both multiple model inference steps for reconfiguration and the waiting time required for system stabilization after applying the recommended parallelisms.}

\begin{table}[]
\caption{Frequency of Backpressure Occurrences}
\vspace{-0.5em}
\centering
\scalebox{0.72}{
\begin{tabular}{|l|cllllcll|}
\hline
\rowcolor[HTML]{C0C0C0} 
\multicolumn{1}{|c|}{\cellcolor{mygrey}} & \multicolumn{8}{c|}{\cellcolor{mygrey}\textbf{Query}} \\ \cline{2-9} 
\rowcolor[HTML]{C0C0C0} 
\multicolumn{1}{|c|}{\multirow{-2}{*}{\cellcolor{mygrey}\textbf{Method}}} & \multicolumn{1}{l|}{\cellcolor{mygrey}Q1} & \multicolumn{1}{l|}{\cellcolor{mygrey}Q2} & \multicolumn{1}{l|}{\cellcolor{mygrey}Q3} & \multicolumn{1}{l|}{\cellcolor{mygrey}Q5} & \multicolumn{1}{l|}{\cellcolor{mygrey}Q8} & \multicolumn{1}{l|}{\cellcolor{mygrey}Linear} & \multicolumn{1}{l|}{\cellcolor{mygrey}2-way-join} & \cellcolor{mygrey}3-way-join \\ \hline
DS2 & \multicolumn{1}{c|}{0} & \multicolumn{1}{c|}{0} & \multicolumn{1}{c|}{1} & \multicolumn{1}{c|}{2} & \multicolumn{1}{c|}{1} & \multicolumn{1}{c|}{3} & \multicolumn{1}{c|}{8} & \multicolumn{1}{c|}{12} \\ \hline
ContTune & \multicolumn{1}{c|}{0} & \multicolumn{1}{c|}{0} & \multicolumn{1}{c|}{2} & \multicolumn{1}{c|}{5} & \multicolumn{1}{c|}{1} & \multicolumn{1}{c|}{4} & \multicolumn{1}{c|}{11} & \multicolumn{1}{c|}{9} \\ \hline
ZeroTune & \multicolumn{1}{c|}{/} & \multicolumn{1}{c|}{/} & \multicolumn{1}{c|}{/} & \multicolumn{1}{c|}{/} & \multicolumn{1}{c|}{/} & \multicolumn{1}{c|}{0} & \multicolumn{1}{c|}{0} & \multicolumn{1}{c|}{0} \\ \hline
\sysname & \multicolumn{1}{c|}{0} & \multicolumn{1}{c|}{0} & \multicolumn{1}{c|}{0} & \multicolumn{1}{c|}{0} & \multicolumn{1}{c|}{0} & \multicolumn{1}{c|}{0} & \multicolumn{1}{c|}{0} & \multicolumn{1}{c|}{0} \\ \hline
\end{tabular}
}
\label{tab:backpressure}
\vspace{-1.5em}
\end{table}

\vspace{-0.5em}
\subsection{Evaluation for Bottleneck Elimination} \label{sec: exp-bottleneck}

We also evaluate the effectiveness of different tuning methods in mitigating bottlenecks by analyzing their ability to eliminate dataflow-level backpressure under varying source rates in Flink. 
Table~\ref{tab:backpressure} reports the total number of backpressure occurrences during different methods' tuning processes throughout the periodic pattern of source rate changes.

Overall, both ZeroTune and \sysname successfully avoid backpressure across all streaming jobs, while DS2 and ContTune trigger backpressure multiple times.
ZeroTune avoids backpressure by consistently recommending higher parallelism degrees to meet its optimized performance objective.
In contrast, DS2 and ContTune rely on the \textit{useful time} metric to estimate operator \textit{processing ability}, which is prone to inaccuracies. 
Overestimation of processing ability leads these methods to recommend insufficient parallelism, resulting in backpressure.
Note, underestimating processing ability does not cause backpressure but results in excessive parallelism and resource wastage.
\sysname addresses these limitations by directly modeling bottleneck occurrence as its prediction objective. This ensures effective bottleneck resolution while avoiding both over- and under-provisioning of parallelism.


\begin{figure}[t]
\centering
\subfloat[\revise{Final parallelism recommendations for Nexmark Queries.}]{
\label{fig: timely_parallelism}
\vspace{-2ex}
{\includegraphics[scale=0.38]{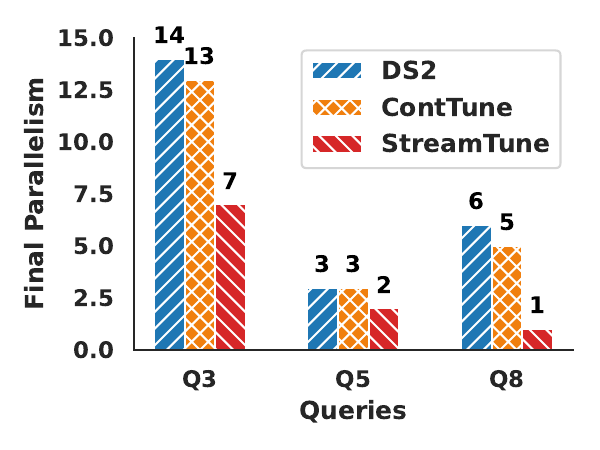}} 
}
\hfill
\subfloat[CDFs of per-epoch latencies for Nexmark Q3.]{
\label{fig: q3_cdf}
\vspace{-2ex}
{\includegraphics[scale=0.38]{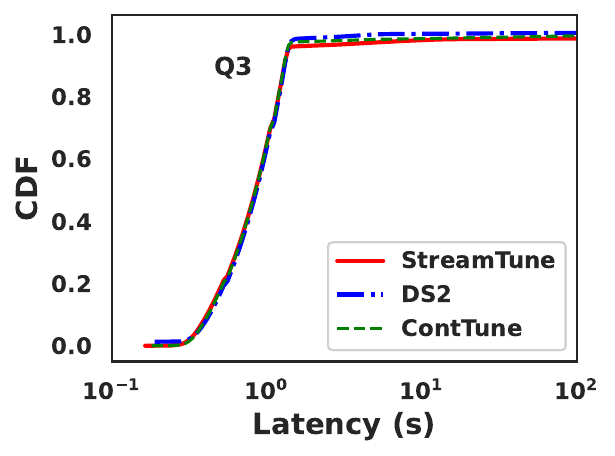}} 
} \ \ \\
\subfloat[CDFs of per-epoch latencies for Nexmark Q5.]{
    \label{fig: q5_cdf}
    \vspace{-2ex}
    {\includegraphics[scale=0.38]{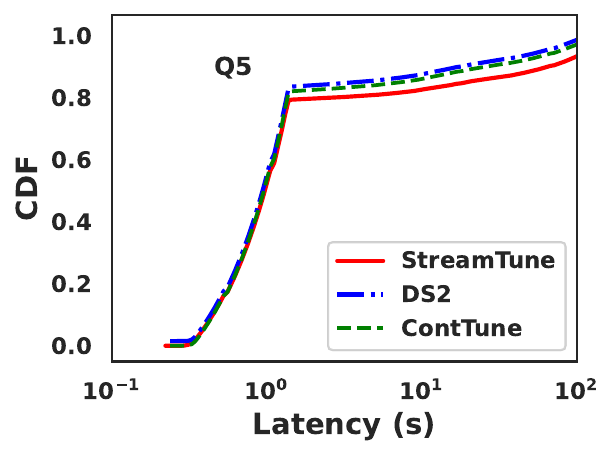}}   
 }
 \hfill
 \subfloat[CDFs of per-epoch latencies for Nexmark Q8.]{
    \label{fig: q8_cdf}
    \vspace{-2ex}
    {\includegraphics[scale=0.38]{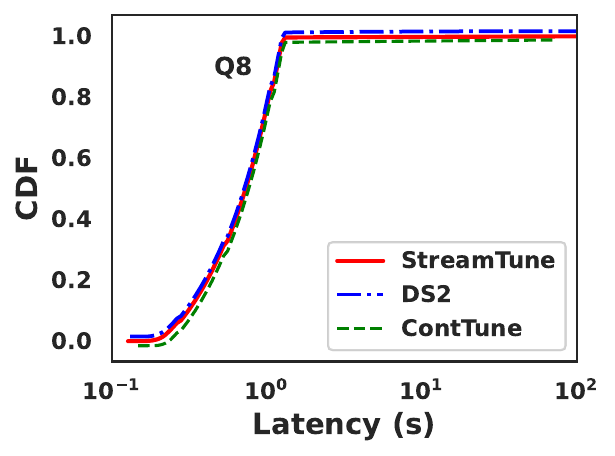}}   
 }
  \vspace{-0.5em}
  \caption{Evaluation Results in Timely Dataflow.}
  \vspace{-2.5em}
  \label{fig:timely}
 
\end{figure}

\vspace{-0.5em}
\subsection{Generality Evaluation on Timely DataFlow} \label{sec: exp_generality}
To demonstrate the generality of \sysname, we conduct a comparative evaluation on Timely Dataflow.
We report results for Q3, Q5, and Q8, as other Nexmark jobs run effectively with a parallelism of 1. Fig.~\ref{fig: timely_parallelism} presents the total operator parallelism recommended by different tuning methods for each streaming job when the source rate changes to $10W_u$.
\sysname consistently recommends lower parallelism than DS2 and ContTune, indicating reduced computational resource requirements. 
The cumulative distribution functions (CDFs) of per-epoch latencies under the parallelism recommendations by different methods are shown in Fig.~\ref{fig: q3_cdf}-\ref{fig: q8_cdf}. 
Note, the per-epoch latency measures the time required to process one epoch of data, where an epoch represents a fixed time interval or a predefined data volume in Timely.
Despite lower parallelism, \sysname matches DS2 and ContTune in query performance, effectively balancing resource efficiency and performance.
For Q8, \sysname needs 83.3\% less parallelism than DS2 while keeping comparable performance.

\begin{figure}[t]
\centering
\subfloat[\revise{\small Online Tuning.}]{
\label{fig: inference_time}
\vspace{-2ex}
{\includegraphics[scale=0.39]{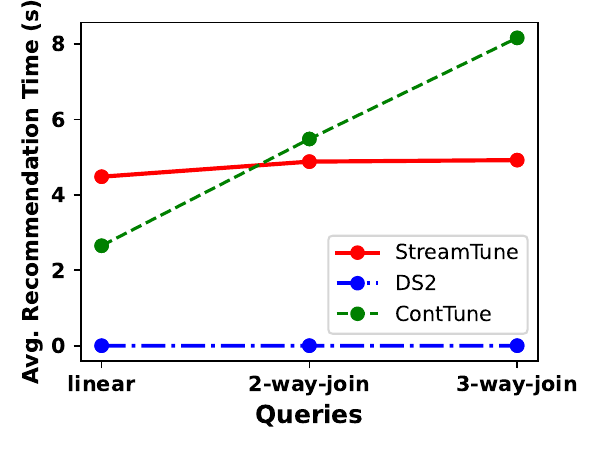}} 
}
\hspace{-1em}
\subfloat[\revise{\small Offline Pre-training.}]{
\label{fig: pretraining_time}
\vspace{-2ex}
{\includegraphics[scale=0.38]{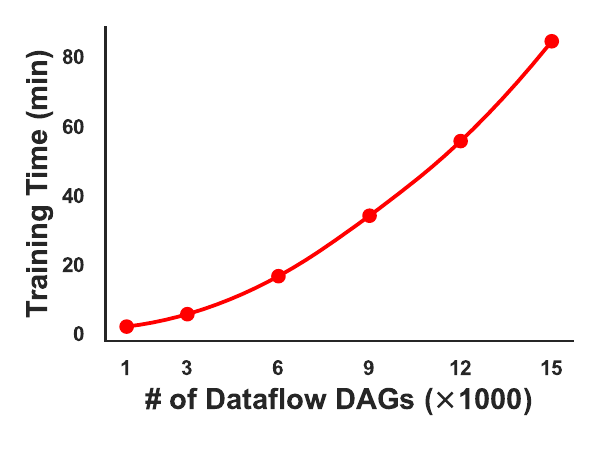}} 
} \ \ \\
\vspace{-1ex}
\caption{\revise{Computational Cost of \sysname.}}
\vspace{-1.5em}
\label{fig: resource_overhead}
\end{figure}

\vspace{-0.5em}
\subsection{\revise{Resource Overhead Evaluation}} \label{sec: exp_overhead}
\revise{
We evaluate the computational cost of \sysname in both online tuning and offline pre-training phases.
In Fig.~\ref{fig: inference_time}, we compare the average recommendation time of \sysname, DS2, and ContTune across different PQP queries.
The results suggest that while \sysname has a relatively longer recommendation time than DS2, it generally outperforms ContTune.
As query complexity increases, \sysname maintains a stable recommendation time, whereas ContTune exhibits a sharp rise.
The evaluation result of pre-training costs is illustrated in Fig.~\ref{fig: pretraining_time}. It demonstrates a non-linear increase as the dataset expands, indicating that pre-training on larger datasets requires significantly more computational resources.}

\begin{figure}[t]
\centering
\subfloat[\revise{\small (Nexmark) Q2.}]{
\label{fig: Q2_CPU}\vspace{-0.5em}
{\includegraphics[trim=0.4cm 0.4cm 0cm 0.4cm, clip, scale=0.3]{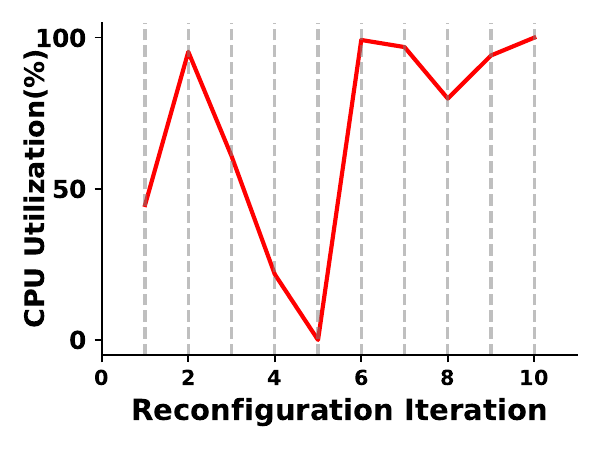}} 
}
\hspace{-0.5cm}
\subfloat[\revise{\small (PQP) Linear.}]{
\label{fig: Linear_CPU}\vspace{-0.5em}
{\includegraphics[trim=0.4cm 0.4cm 0cm 0.4cm, clip, scale=0.3]{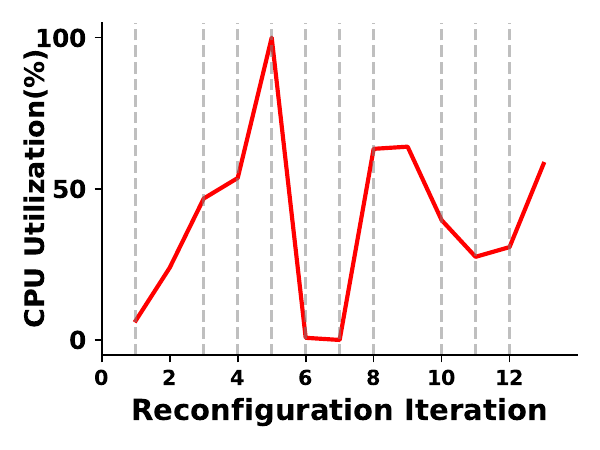}} 
}
\hspace{-0.5cm}
\subfloat[\revise{\small (PQP) 2-way-join.}]{
    \label{fig: 2WayJoin_CPU}\vspace{-0.5em}
    {\includegraphics[trim=0.4cm 0.4cm 0cm 0.4cm, clip, scale=0.3]{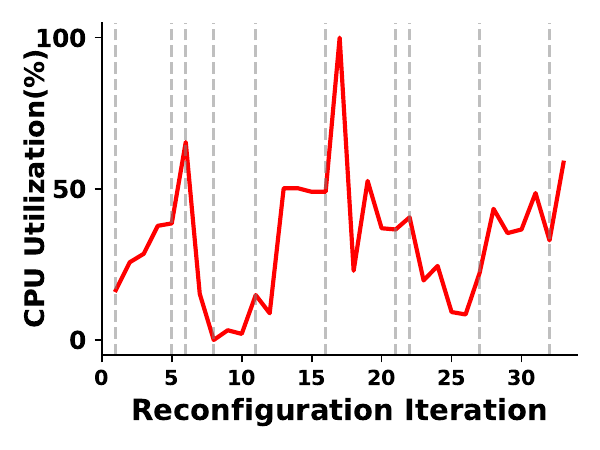}}   
 }  
 \vspace{-0.5em}
  \caption{\revise{CPU Utilization Trends During Tuning Process.}}
  \vspace{-2em}
  \label{fig:cpu_utilization}
 
\end{figure}

\vspace{-0.5em}
\subsection{\revise{CPU Utilization Evaluation During Reconfiguration}} \label{sec:cpu_utilization}
\revise{
Fig.~\ref{fig:cpu_utilization} illustrates the fluctuation in CPU utilization over multiple reconfiguration iterations while \sysname tunes parallelism on Apache Flink for different streaming jobs.
The vertical dotted line indicates the periodic change in source rates.
The observed trends highlight the dynamic adjustments in resource usage as \sysname iteratively conducts tuning.
Notably, CPU utilization exhibits significant variations across reconfiguration iterations, suggesting that \sysname actively explores different parallelism degrees.
This behavior is 
evident in more complex queries, where a higher number of reconfigurations are required to adapt to different source rates.}

\subsection{Ablation Study} \label{sec: exp_abalation}

\begin{figure}[t]
\centering
\hspace{-1cm}
\subfloat[{\small Effect of Classification Models.}]{
\label{fig: model_reconfigure}
\vspace{-1.5ex}
{\includegraphics[trim=0.4cm 0.4cm 0cm 0.4cm, clip, scale=0.35]{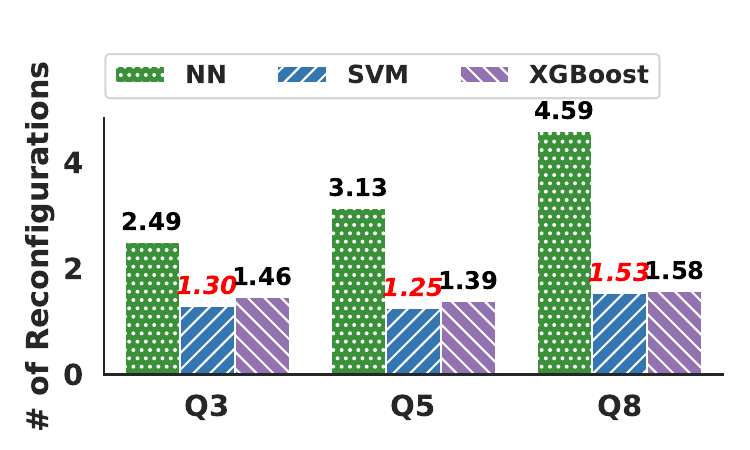}} 
} 
\subfloat[{\small Processing Time.}]{
\label{fig: ged_computing_cost}
\vspace{-1.5ex}
{\includegraphics[trim=0.4cm 0.4cm 0cm 0.4cm, clip, scale=0.25]
{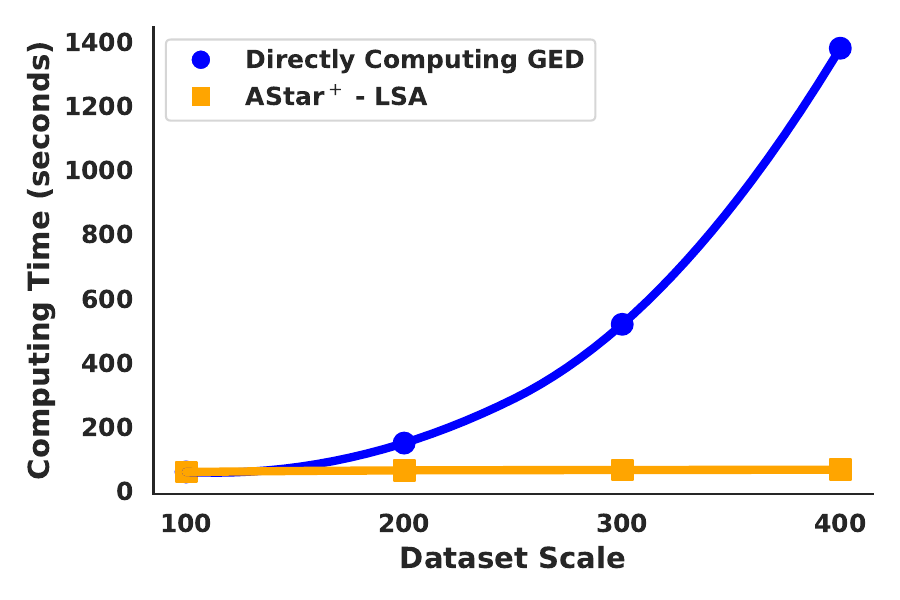}} 
}
\ \ \\
\vspace{-0.5em}
\caption{{Ablation Study.}}
\vspace{-1em}
\end{figure}

\noindent \textbf{Effect of Classification Models.} 
We analyze the impact of the prediction layer on tuning effectiveness with three Nexmark queries. Three models are evaluated: SVM and XGBoost, which satisfy the monotonic constraint, and Neural Network (NN), which does not. 
%
The evaluation of each model's ability to eliminate backpressure in Flink is shown in Fig.~\ref{fig: model_reconfigure}.
Overall, SVM and XGBoost perform comparably, while NN fails to provide optimal parallelism that effectively minimizes backpressure occurrences.
For Q3 and Q5, although NN recommends lower parallelism degrees, it incurs higher backpressure occurrences. 
This is because NN does not enforce the monotonic constraint, which allows for consistent increases in an operator’s processing ability through parallelism. 

\noindent \textbf{Processing Time of Similarity Center Computation.}
We examine the processing time for computing the similarity center.
%
%
%
%
The processing time for similarity center computation depends 
on the efficiency of the graph similarity search. We compare two methods: directly computing GED and using the AStar$^+$-LSa algorithm.
Figure~\ref{fig: ged_computing_cost} presents the results. As the number of clustering DAGs increases, the time required for direct GED computation grows significantly. In contrast,  AStar$^+$-LSa maintains consistently low processing times. Notably, for a dataset with $400$ dataflow DAGs, it reduces the time cost by $99.65\%$ compared to direct GED computation.

\section{Related Work}

\noindent \textbf{Parallelism tuning for streaming systems.} Dhalion~\cite{floratou2017dhalion}, a system developed for the Heron stream processing framework, employs some heuristics to optimize job parallelism. However, it lacks the flexibility to adapt to diverse workloads.
DS2~\cite{kalavri2018three} is a more analytical approach. It operates under the assumption of a linear relationship between parallelism and processing ability. The system aims to recommend the lowest degree of parallelism that satisfies processing ability requirements, thereby optimizing resource utilization. However, the linear relationship may not hold in all scenarios, potentially limiting its applicability in many jobs.
Addressing some of the limitations of DS2, ContTune~\cite{lian2023conttune} leverages Bayesian optimization to model the relationship between parallelism and processing ability for different operators within jobs. Then it applies a big-small algorithm for tuning. Although this approach offers a more nuanced understanding of parallelism impacts, it does not exploit the structural information inherent in streaming jobs, and its ability to utilize historical data remains limited.
ZeroTune~\cite{agnihotri2024zerotune} incorporates GNN to develop a cost model for parallelism tuning. This method shows promise in capturing interactions within stream processing systems. However, the integration of this cost model into parallelism optimization remains a challenge, limiting its practical applicability.


\noindent \textbf{Configuration tuning for data management systems.}  It is important to recognize that stream processing systems and other data management systems have a wide array of configurable parameters beyond just parallelism. The application of machine learning techniques to optimize system performance across these various parameters has emerged as a significant area of research in recent years.
GML~\cite{guo2021gml} represents a notable advancement in this field, proposing a guided machine learning approach specifically tailored to optimize the overall performance of Flink. 
OtterTune~\cite{van2017automatic}, while not specifically designed for streaming systems, has made significant contributions to the broader field of database system tuning. It introduces a Lasso-based knob selection method coupled with a Gaussian Process Regression-based tuning paradigm. 
Building on these foundations, ResTune~\cite{zhang2021restune} introduces an innovative approach that extracts meta-features from workloads and applies them to database tuning. This method has shown promise in accelerating tuning processes, potentially reducing the time and resources required for effective system optimization. 
CDBTune~\cite{zhang2019cdbtune} takes yet another approach, leveraging reinforcement learning to optimize database configuration. 
\revise{
\section{Discussions and Future Work}
\noindent \textbf{Limited Pre-training Dataset.}
When the dataset for pre-training is limited, the effectiveness of \sysname may be diminished. To handle this situation, \sysname can bypass the GED similarity-based clustering process and 
train a global GNN encoder. During the online tuning phase, fine-tuning can be conducted directly with this global encoder, eliminating the need for the similarity-based search of a target encoder.

\noindent \textbf{Unseen Streaming Workloads.}
\sysname generalizes the tuning process by leveraging structural clustering, GNN-based encoding of dataflow DAGs, and online tuning for continuous adaptation.
However, like ZeroTune, it typically employs one-hot encoding for operator features (e.g., operator types), 
requiring retraining when entirely new operators are introduced.
Future work could explore embedding-based representations that capture semantic relationships between operators, improving generalization to unseen operators. 

\noindent \textbf{Live Reconfiguration.} One limitation of \sysname is its reliance on stop-and-restart reconfiguration, which incurs downtime. Future work could explore live reconfiguration mechanisms, such as operator-level RESTful APIs, to dynamically adjust parallelism without restarting the system. Such an approach has been deployed internally at ByteDance, where operators are assigned parallelism dynamically through APIs, enabling the Flink JobManager to apply changes at runtime.

\noindent \textbf{Scheduling-Aware Tuning.}
\sysname can be extended to incorporate scheduling-aware tuning, particularly for those DSPSs lacking built-in load balancing and robust resource management like Timely Dataflow.
Enhancing resource-aware tuning could improve adaptability in environments where dynamic scheduling mechanisms are not available.
}

\section{Conclusion}

In this paper, we propose \sysname, an adaptive parallelism tuning approach for stream processing systems. \sysname employs a pre-training and fine-tuning framework to balance resource usage and processing performance. 
Its GNN-based encoding and clustering techniques enable efficient and adaptable tuning across diverse workloads.
Experimental results show that \sysname outperforms state-of-the-art methods by reducing operator parallelism, eliminating backpressure, and minimizing reconfigurations. 

\balance

\clearpage
\bibliographystyle{IEEEtran}
\bibliography{IEEEabrv,ref}

\end{document}